\documentclass[12pt]{article}

\usepackage{subfigure}
\usepackage{graphicx}
\usepackage{url}
\usepackage{amsmath}
\usepackage{multicol}
\usepackage{fancyvrb}

\begin{document}

\title{An Evaluation of Caching Policies for Memento TimeMaps}


\author{
Justin F. Brunelle \& Michael L. Nelson\\
       Old Dominion University\\
       Department of Computer Science\\
       Norfolk, Virginia, 23508\\
       \{jbrunelle, mln\}@cs.odu.edu
}

\maketitle
\begin{abstract}
As defined by the Memento Framework, TimeMaps are ma-chine-readable
lists of time-specific copies -- called ``mementos'' -- of an archived original resource.
In theory, as an archive acquires additional mementos over time, a
TimeMap should be monotonically increasing.
However, there are reasons why the number of mementos
in a TimeMap would decrease, for example: archival redaction of
some or all of the mementos, archival
restructuring, and transient errors on the part of one or more
archives.  We study TimeMaps for 4,000 original resources over a
three month period, note their change patterns, and develop a caching algorithm for
TimeMaps suitable for a reverse proxy in front of a Memento aggregator.
We show that TimeMap cardinality is constant or monotonically increasing for 80.2\% of all TimeMap downloads observed in the observation period.
The goal of the caching algorithm is to exploit the ideally monotonically
increasing nature of TimeMaps and not cache responses with fewer
mementos than the already cached TimeMap.  
This new caching algorithm uses conditional cache replacement and a Time To Live (TTL) value to ensure the user has access to the most complete TimeMap available. Based on our empirical data, a TTL of 15 days will minimize the number of mementos missed by users, and minimize the load on archives contributing to TimeMaps.
\end{abstract}




\section{Introduction}
\label{introduction}
The Memento Framework provides HTTP extensions for inter-archive communication and integrating the past and current Web \cite{nelson:memento:tr, memento:rfc, ldow2010:memento}. Memento TimeMaps, part of the Memento Framework, provide an aggregate view of mementos of a URI-R existing in distributed archives as a single document called a TimeMap. These TimeMaps, identified by a URI-T, contain a set of URI-Ms that each have a datetime on which they were archived, or a Memento-Datetime \cite{nelson2011memento}. For example, the TimeMap shown in Figure \ref{tm} gives a set of mementos aggregated from two repositories, the National Archives of UK and the Internet Archive's Wayback Machine.

\sloppy
 It also spans from the first memento at Dec. 12, 2007 to the last memento at Dec. 14, 2011. Because TimeMaps can aggregate lists of URI-Ms from several sources, many factors can influence the cardinality of a TimeMap, including network downtime, availability of the archive, hardware malfunctions, routine maintenance, etc. Human-induced interruptions in memento availability can also occur. For example, the robots.txt protocol is a redaction method at the Internet Archive (IA) that has been well documented\footnote{\url{http://archive.org/post/778/exclusions-from-the-wayback-machine}}. Copyright and content sensitivity can also cause mementos to be taken out of public access \cite{copyright, copyright2}. These redaction requests are legitimate removals of mementos at the content owners' request, and not a failure on the part of the archives\footnote{\url{http://www2.sims.berkeley.edu/research/conferences/aps/removal-policy.html}}.
\fussy

\sloppy
A related modification of archives' offerings is URI migration. For example, the migration of a URI such as \url{http://example.org/archives/http://thesite.com/} to a URI of \url{http://memento.example.org/archive/http://thesite.com/} would change all URI-Ms for the archive's contributions to the TimeMap of URI-R. Migrations of this nature assign new URIs to existing mementos.
\fussy

\begin{figure*}[ht!]
{\footnotesize
  \begin{verbatim}
<http://http://mementoproxy.cs.odu.edu/aggr/Timemap/link/http://flare.prefuse.org/>;
   		rel="self"; type="application/link-format",
<http://mementoproxy.cs.odu.edu/aggr/timegate/http://flare.prefuse.org/>;
   		rel="timegate",
<http://flare.prefuse.org/>;rel="original",
<http://api.wayback.archive.org/memento/20071213002102/http://flare.prefuse.org/>;
   		rel="first memento"; datetime="Thu, 13 Dec 2007 00:21:02 GMT",
<http://api.wayback.archive.org/memento/20080509125659/http://flare.prefuse.org/>;
   		rel="memento"; datetime="Fri, 09 May 2008 12:56:59 GMT",
<http://webarchive.nationalarchives.gov.uk/20080908074106/http://flare.prefuse.org/>;
   		rel="memento"; datetime="Mon, 08 Sep 2008 00:00:00 GMT",
   ...
<http://api.wayback.archive.org/memento/20100815085828/http://flare.prefuse.org/>;
   		rel="memento"; datetime="Sun, 15 Aug 2010 08:58:28 GMT",
<http://webarchive.nationalarchives.gov.uk/20100909131056/http://flare.prefuse.org/>;
  		rel="memento"; datetime="Thu, 09 Sep 2010 00:00:00 GMT",
<http://api.wayback.archive.org/memento/20101107020354/http://flare.prefuse.org/>;
   		rel="memento"; datetime="Sun, 07 Nov 2010 02:03:54 GMT",
  \end{verbatim}
}
  \caption{Example Partial TimeMap for URI-R \protect\url{http://flare.prefuse.org/}.}
  \label{tm}
\end{figure*}

\subsection{Caching TimeMaps}
Since TimeMaps do not always improve when they change (e.g., due to archive unavailability), caching policies become important -- traditional caches could potentially cache a worse TimeMap than should be available.
TimeMaps ideally follow a monotonically increasing growth pattern. That is, TimeMaps should never \emph{lose} mementos (except in rare cases of redaction) -- mementos should remain listed in a TimeMap after their first appearance. Since TimeMaps are expensive to generate and change slowly, they are good candidates for caching. 

\subsection{Contributions}
This paper studies a set of 4,000 URI-Rs; 1,000 URI-Rs each came from the Open Directory Project, Delicious, Bit.ly, and search engine caches \cite{hmotwia}. We observed the TimeMaps of the 4,000 URI-Rs for a 3-month period (from May 1st to July 31st, 2012 for a total of 92 days) and we mapped and evaluated their evolution and change patterns. This paper shows that TimeMap cardinality is monotonically increasing (i.e., the same or increasing) for 80.2\% of all TimeMap changes in the observation period.

We also proposes a new caching algorithm utilizing a TTL value for caching Memento TimeMaps. This caching algorithm exploits the knowledge that TimeMaps sometimes do not change ``for the better'' and therefore should not always be cached. The change patterns of the observed TimeMaps are used to empirically determine the optimal TTL value.

\section{Related Work}
\label{priorwork}

It is important to understand the behavior of TimeMaps and the associated availability of mementos. For example, the Warrick project \cite{frank_lazy} uses TimeMaps to recover mementos with the ultimate goal of recovering lost websites. Obviously, understanding how archives advertise their mementos and understanding how the availability changes is important for several aspects of Warrick. Caching TimeMaps is used to balance load on the archives vs. Warrick's need for the most recent mementos.

Due to the popularity of Memento, other services and tools have begun to consume and rely on TimeMaps to provide functionality.  For example, the UK Web Archive is providing a visualization tool for TimeMaps \cite{mementoviz}. Additionally, Android's Memento browser \cite{sandersonimplementing} and an iOS mobile memento browser \cite{franktbd} all rely on TimeMaps to provide time travel for the Web.

Observing and studying change on the Web is not new; several works have observed the changes that Web pages undergo over time, proving their ephemeral nature \cite{cho2000evolution, 1267293, brewington2000keeping, fetterly2004large, ntoulas2004s, dl_persistence}. In contrast to these studies, TimeMaps can be thought of as Web resources that cannot be evaluated as ``fresh'' solely on their age or change status.

Disk failures are one reason for changes in memento availability. A case study of failures at the Internet Archive (IA) is provided in Schwarz's 2006 work \cite{Schwarz2006}. The organization of archives is also important; we must understand how an entire server failure affects the availability of mementos. Jaffe's work in 2004 describes the IA's architecture \cite{Jaffe:2009:AIA:1534530.1534545}.

Caching is an important addition to the Web. It reduces latency, load, and user wait times. A set of caching methods commonly utilized on the Internet -- including TTL values -- are discussed in Wang's 1999 survey paper \cite{webCaching}. Other work has been done to benchmark the performance boosts achieved when implementing caching of dynamic contents \cite{cacheDynamic}. Other works have studied best efforts for caching dynamically generated content, such as that generated on the server-side \cite{activeCache, classcache}.

Reverse proxies such as Squid \cite{wessels2004squid} improve the efficiency of Web traffic by caching content. This caching is independent of the content server and requester, allowing both to operate without modification. The work we present can be implemented using a reverse proxy (e.g., Squid allows custom caching rules) or by modifying the Memento aggregator cache, and will reduce the load on the Memento proxies while increasing their reliability for users. 

\section{Experiment design}

\label{experiment}
We observed TimeMap cardinality (number of mementos present in a TimeMap) daily for 4,000*92=386,400 total download attempts and the consistency of URI-Ms and the datetime values associated with a memento throughout the experiment. Once a day, a set of scripts downloaded the 4,000 TimeMaps so that we can analyze the daily changes of the mementos advertised in the TimeMaps, as well as how the contents of the TimeMaps differ between observations. To ensure the 4,000 TimeMaps could be retrieived in a 24 hour period, the scripts used a timeout for accessing the TimeMaps. A 45 second timeout was used because this is the upper limit that humans are willing to wait when accessing a resource \cite{nah}. The experiment also ran 11 parallel scripts to access different portions of the 4,000 URI dataset. With this concurrency and limited wait time, all 4,000 TimeMaps were accessed consistently within a 24 hour period. 

\subsection{Cache-less Memento Proxies}
\label{proxies}
The first step in the experiment setup was to create and install a new set of Memento proxies separate from the production proxies at Los Alamos National Laboratory and Old Dominion University. The existing proxies are constantly being accessed by MementoFox, users, and other experiments. We installed a set of proxies on a separate, private experiment machine (\texttt{128.82.7.240}) to prevent any contamination from the public Memento uses. The proxies were installed just as they exist on the production Memento machines. However, the production proxies cache their responses in an infinitely-sized cache with a TTL=$\infty$  to speed up the response time to the users, as well as to prevent unnecessary load on the archives. The caching software had to be removed from the experiment proxies to ensure fresh results of each observation. These memento proxies queried for, constructed, and served the TimeMaps throughout the experiment run.

\subsection{TimeMaps as Web Resources}
\label{timemapResources}
TimeMaps can change, evolve, and grow like traditional Web resources. However, TimeMaps differ from traditional Web resources in that they do not always change ``for the better.'' Traditional Web resources can be maintained in a cache by monitoring any change to the resource content. TimeMaps can change in a detrimental way -- they sometimes lose mementos due to intermittent contributions from archives, URI changes, or archival redactions. Archival redactions will results in a HTTP 404 response when dereferencing a URI-M but redactions are very rare. We are more concerned about transient errors. As such, TimeMaps cannot be cached in the same manner -- only TimeMaps that have been changed in a positive way should be updated in a cache. Prior to this work, there was no method to designate a positive or negative change to a TimeMap.

The production Memento proxies utilize caching to limit the load on the archives and increase response time to users. The proxies are research prototypes, and as such, the caching algorithm for Memento was designed around one of two simple solutions. The two simplest caching algorithms are to cache everything or cache nothing. Caching nothing induces unneeded and unfair load on the archives contributing to the TimeMaps and increases the service time for users due to the latency between the Memento Proxies and archives. Therefore, the cache everything approach was taken. The proxies cache the first TimeMap observed in system and holds it until a cleansing of the cache is performed manually. If a \emph{bad} TimeMap is cached, it persists in the cache until the cache is manually cleaned. TimeMaps that are entirely empty (i.e., receive an HTTP 404 response) are especially bad (the TimeMaps show that a URI-R is not archived, when in fact, it may have mementos that were not reported), and are discussed in more depth in Section \ref{caseSection}.

\subsection{Experiment Execution}
\label{execution}
\begin{figure}[h!]
    \includegraphics[width=1\textwidth]{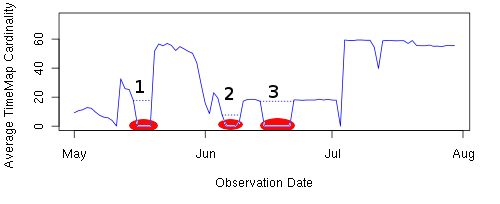}
  \caption{Average TimeMap Cardinality during experiment execution (2012).}
\label{optavg}
\end{figure}

During the experiment, three outages were observed as noted by the red circles in Figure \ref{optavg}. Annotation 1 in the figure indicates where the locally installed proxies were inoperable from May 16 -- May 18 due to a system reset of the department machines. As indicated by 2 in the figure, the Internet Archive proxies were inoperable due to edits to the API occurring at the Internet Archive. As indicated by 3 in the figure, there was a massive power failure at at our university that caused the machines to automatically reboot, killing the experiment run. This failure went undetected for 6 days in June.

We took these time periods of low memento availability into consideration during the calculations of our results. The previous TimeMaps were substituted for the missing TimeMaps, effectively treating them as unchanged instead of non-existent. 

\section{Experiment Results}
\label{results}
It was immediately clear that TimeMaps were not monotonically increasing. That is, TimeMaps sometimes get smaller -- or ``worse'' instead of staying the same or growing. 

\subsection{TimeMap Change Types}
\label{caseSection}

Utilizing the collected TimeMaps over the course of the observation period, we categorized the changes to the TimeMaps. We used the classifications in Table \ref{classTable} to determine how changes affected the TimeMaps. The table uses \emph{a} to denote the number of archives that contribute to a TimeMap at time \emph{t} and \emph{a'} to denote the number of archives that contribute to a TimeMap at time \emph{t+1}. Similarly, the table uses \emph{m}  to denote the number of mementos in the TimeMap at time \emph{t} and \emph{m'} to denote the number of mementos in the TimeMap at time \emph{t+1}.

\begin{figure}[h!]
    \includegraphics[width=1\textwidth]{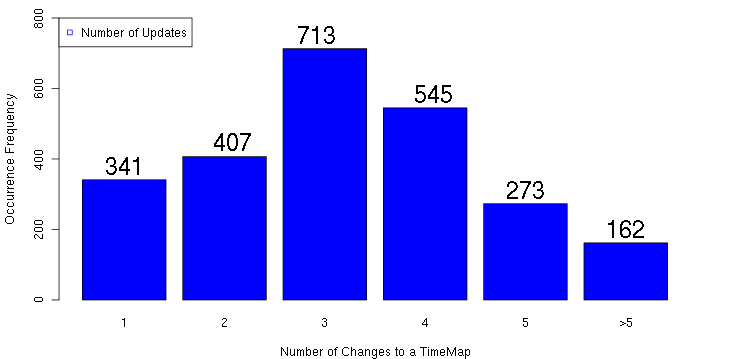}
  \caption{Number of changes to TimeMaps.}
\label{optimalttls}
\end{figure}

TimeMaps can lose and gain mementos and contributing archives. This experiment utilizes the change patterns of TimeMaps to predetermine the change rates for the observed URI-Rs. The number observed changes are provided in Figure \ref{optimalttls}. Most TimeMaps changed 3 or 4 times over the course of the experiment.

As shown in Figure \ref{optimalttls}, most TimeMaps underwent 3 changes throughout the observation period. Observing these change rates provides insight into the most appropriate TTL values for the TimeMaps. It also disregards TimeMaps that were always empty during the collection part of the experiment. On average, the TimeMaps observed in this experiment changed every 37.6 days, with the most frequently occurring change rate observed of every 30 days, which roughly equates to 3 changes throughout the observation period. 

Mementos can be added by contributing archives, or mementos can be redacted by contributing archives. Contributing archives can also disappear from a TimeMap due to service interruption or other unavailability. New archives can begin contributing to a TimeMap if a new URI-R is added to the collection. Cases 6 and 7 are the most detrimental to a TimeMap -- the overall cardinality of the TimeMap is reduced by the loss of mementos. Case 1 is the most observed case, and represents the TimeMap remaining consistent between two times. Cases 2, 3, 4, and 5 all result in additional mementos being added and are therefore improvements upon the TimeMap at the previous time. Case 4 is unique in that an archive is lost but there are still new mementos in the TimeMap; we lean toward updating the cache in Case 4. Given that Case 6 is the most frequently observed change to TimeMaps, we show that if a TimeMap changes, it is most often for the worse. The occurrence of cases throughout the experiment is provided in Table \ref{classTable}. These classifications of TimeMaps establish the notion of \emph{better} and therefore how to handle cache replacement.

\begin{table*}[t]
\includegraphics[width=1\textwidth]{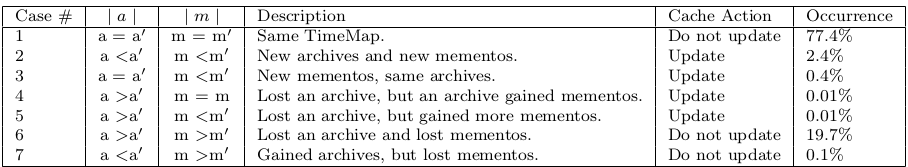}
\caption{Classifications of TimeMap Changes}
\label{classTable}
\end{table*}

\subsection{Definition of TimeMap Changes}
\label{strictvLoose}
The cardinality of a TimeMap at observation time \emph{t} is shown as the red lines in Figures \ref{firstGraph1} - \ref{comp4}. The blue lines are the cumulative size of the set of all unique URI-Ms up to time \emph{t}: [0, \emph{t}]. 

A memento is defined in Equation \ref{mem} as  a version of a URI-R at time \emph{k}.

\begin{eqnarray}
\label{mem}
m_{R, k} = &(URI-R, k), \\& k \in \{\mbox{times in RFC 1123 format \cite{rfc1123}}\} \nonumber
\end{eqnarray}

The URI that identifies a memento is defined in Equation \ref{urim}.

\begin{equation}
\label{urim}
URI-M_{R, k} = \mbox{URI of } m_{R, k}
\end{equation}

TimeMaps are composed of mementos of a URI-R, as defined in Equation \ref{TM}.

\begin{equation}
\label{TM}
TM_{R, t} = \bigcup_k URI-M_{R, k}
\end{equation}

The cardinality of a TimeMap $TM_{R, t}$ is defined in Equation \ref{card}. Note that  this refers to the number of unique URI-Ms that appeared in a URI-R's TimeMap as observed at time \emph{t}. 

\begin{equation}
\label{card}
\left|TM_{R, t}\right|
\end{equation}

Thus, monotonically increasing TimeMaps would satisfy the condition in Equation \ref{min}.

\begin{equation}
\label{min}
\left|TM_{R, t_1}\right| \leq \left|TM_{R, t_2}\right|, \mbox{where } t_1 < t_2
\end{equation}

\subsection{Strict versus Loose Policy}
\label{strict}
When comparing URI-Ms in a TimeMap to determine how TimeMaps grow over time, lexigraphically comparing URI-Ms is the intuitive method. We will refer to this method of matching as the \emph{Strict Policy}. 

\sloppy
The best possible set of URI-Ms is the cumulative set of all URI-Ms that have been observed in the TimeMaps until a time \emph{t}. This represents the best possible TimeMap we would see if TimeMaps were monotonically increasing. This is represented by Equation \ref{strictEqn}.
\fussy

\begin{equation}
\label{strictEqn}
M_{strict} =  \bigcup_t TM_{R, t}
\end{equation}

URIs are not expected to change \cite{coolUris}, but, due to archives being restructured or changes in the URI-M structure, new URIs identifying existing mementos sometimes occur. The \emph{Loose Policy} only matches the tuple of Memento-Datetime and archive ($archive(URI-M_{R, k})$, k) of a URI-M and the URI-R for which it is a memento. The archive is determined by the hostname and occasionally path of the URI-M. The \emph{Loose Policy} is immune to URI changes and should recognize mementos identified by different URI-Ms across time, testing the theory that when mementos disappear from TimeMaps, they actually just change URI-Ms, and not Memento-Datetimes. Conditions under which this occurs include architecture modifications by the archives, URI scheme changes, or even errors with the URI-M listed.

TimeMap cardinality under the \emph{Loose Policy} is measured in accordance with Equation \ref{looseEqn}.

\begin{eqnarray}
\label{looseEqn}
M_{strict}=&\bigcup_t TM_{R, t} \nonumber\\ & \forall  \text{unique values of }archive(URI-M_{R, k}) \text{ and } k
\end{eqnarray}

For example, the TimeMap in Figure \ref{tmkendrick}, the URI-Ms would \emph{Loosely} match the other URI-Ms with the same Memento-Datetimes. Since they have the same archive (\url{web.archive.org}) and Memento-Datetime (\texttt{Mon, 01 Nov 2010 06:02:04 GMT}). All would be considered lexigraphically different and not match according to the \emph{Strict Policy} because of the same Memento-Datetime but different URI-R (\url{http://aarp.org:80/Health/}).

\begin{figure*}[ht!]
{\footnotesize
  \begin{verbatim}
<http://web.archive.org/web/20101101060204/http://aarp.org:80/Health/>;
	rel="memento";datetime="Mon, 01 Nov 2010 06:02:04 GMT",
<http://web.archive.org/web/20101101060204/http://www.aarp.org:80/Health/>;
	rel="memento";datetime="Mon, 01 Nov 2010 06:02:04 GMT",
<http://web.archive.org/web/20101101060204/http://www.aarp.org:80/health/>;
	rel="memento";datetime="Mon, 01 Nov 2010 06:02:04 GMT",
\end{verbatim}
}
  \caption{A subset of an example TimeMap for URI-R \protect\url{http://aarp.org/Health/}.}
  \label{tmkendrick}
\end{figure*}

Comparisons of cumulative mementos under the \emph{Loose} and \emph{Strict Policies} show that mementos receive new $URI-M_{r, k}$ but are still the same memento $m_{r, k}$. The \emph{Loose Policy} graphs do not show an increase when this activity takes place, while the \emph{Strict Policy} mistakes this new $URI-M_{r_k}$ as a new memento. 

\begin{figure*}[ht!]
  \begin{center}
  \subfigure[\emph{Loose Policy} for yardsellr.com]{\label{loose}\includegraphics[width=1\textwidth]{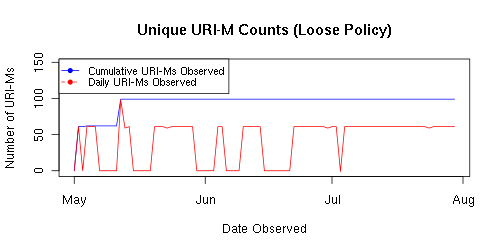}} 
   \subfigure[\emph{Strict Policy} for yardsellr.com]{\label{strict}\includegraphics[width=1\textwidth]{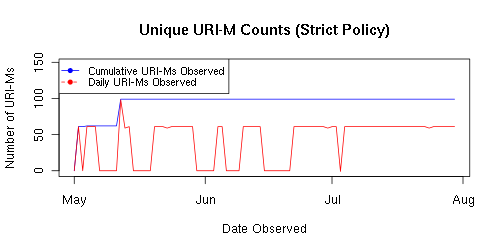}}
  \end{center}
  \caption{The TimeMap of \protect\url{http://yardsellr.com} shows identical graphs for \emph{Strict} and \emph{Loose Policies}, and demonstrates that mementos blink on and off in the TimeMap, but do not change the overall cumulative set of all observed mementos.}
\label{firstGraph1}
\end{figure*}

\begin{figure*}[ht!]
  \begin{center}
  \subfigure[\emph{Strict Policy} for whitehouse.gov]{\label{strict3}\includegraphics[width=1\textwidth]{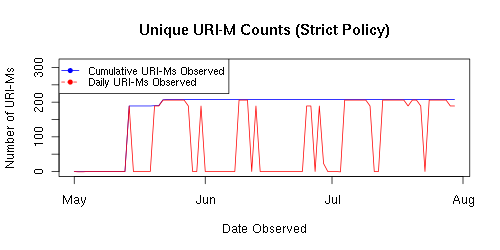}}
    \subfigure[\emph{Loose Policy} for whitehouse.gov]{\label{loose3}\includegraphics[width=1\textwidth]{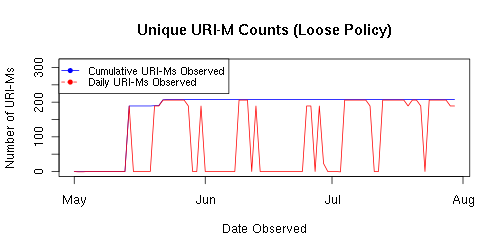}} 
  \end{center}
  \caption{The \protect\url{http://www.whitehouse.gov/administration/eopcea/} TimeMap is well-behaved, with identical \emph{Strict} and \emph{Loose Policy} graphs.}
\label{firstGraph}
\end{figure*}

The two graphs of the yardsellr.com TimeMap, (Figures \ref{strict} and \ref{loose}) and whitehouse.gov TimeMap, (Figures \ref{strict3} and \ref{loose3}) both demonstrate well-behaved TimeMaps in which the \emph{Loose} and \emph{Strict Policy} graphs match one another. This shows that there is no difference between the TimeMap cardinalities according to the \emph{Loose} and \emph{Strict Policies}. However, these two TimeMaps change in different ways during the observation period. The yardsellr.com TimeMap incurs a spike of mementos in mid-May due to an influx of mementos from the Internet Archive. These mementos do not reappear in the TimeMap for the duration of the observation period, which results in a gap between the cumulative and observed mementos (observed as the gap between the blue and red lines, respectively). Alternatively, the whitehouse.gov TimeMap frequently has observed cardinality equal to the cumulative cardinality, thus showing that the maximum number of mementos frequently appears in the TimeMap. The consistent whitehouse.gov TimeMap demonstrates expected behavior from a TimeMap. There are some slight dips in TimeMap cardinality caused by an archive being temporarily unavailable. These dips are not permanent, as seen in the yardsellr.com TimeMap.

\begin{figure*}[h!]
  \begin{center}
  \subfigure[\emph{Strict Policy} for Google Code API]{\label{strictdip}\includegraphics[width=1\textwidth]{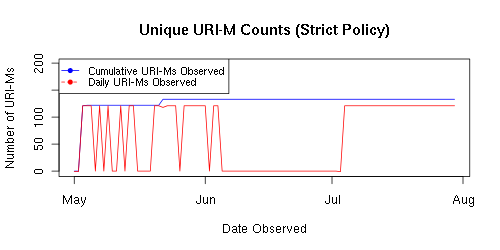}}    
  \subfigure[\emph{Loose Policy} for Google Code API]{\label{loosedip}\includegraphics[width=1\textwidth]{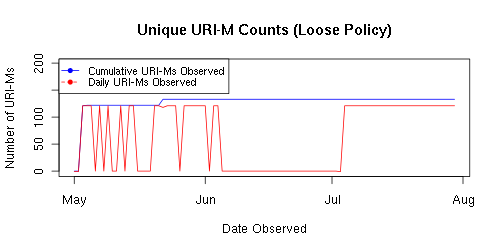}} 
  \end{center}
  \caption{Poorly formed Memento-Datetimes appear in the TimeMaps for \protect\url{http://code.google.com/apis/g/data/client-cs.html}}
\label{comp4}
\end{figure*}

An additional scenario can occur when a memento, or subset of mementos, disappears from the TimeMap altogether. This is observed by the dip representing the loss of URI-Ms from the Google Code API TimeMap on May 22nd in Figures \ref{strictdip} and \ref{loosedip}. This is noteworthy because there is a complete swapping of mementos that occurs. While the observed cardinality decreases, there is an increase in the cumulative number of mementos (the red line goes down while the blue line goes up). This is due to the swapping of an entire set of URI-Ms. 

\begin{figure*}[h!]
  \begin{center}
   \subfigure[\emph{Strict Policy} for Google Translate]{\label{strict4}\includegraphics[width=1\textwidth]{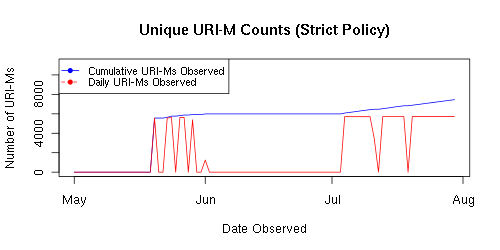}}
   \subfigure[\emph{Loose Policy} for Google Translate]{\label{loose4}\includegraphics[width=1\textwidth]{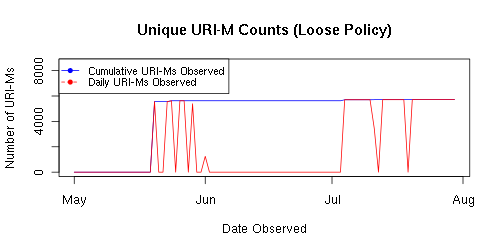}}
  \end{center}
  \caption{TimeMap with duplicate Memento-Datetime values as shown by the \emph{Strict} and \emph{Loose Policies} for \protect\url{http://translate.google.com/}.}
\label{comp41}
\end{figure*}

Google Translate has a TimeMap cardinality $\left|TM\right|$ = 5,800. These mementos are populated from several different archives, but the majority of URI-Ms come from the Internet Archive and from Archeif Web. The TimeMap size over time is provided in Figures \ref{strict4} and \ref{loose4} using the \emph{Strict} and \emph{Loose Policies}, respectively. These graphs differ greatly in their representation of the TimeMap sizes.

When using the \emph{Strict Policy}, the TimeMap appears to be nearly consistently growing in cardinality (the blue line, and number of total unique URI-Ms observed over the observation period), while showing only slight (nearly static) growth in the cardinality of the TimeMaps observed on a given day. This is due to URI-Ms with the same Memento-Datetime appearing in a TimeMap for a short span of time and never returning. 

The \emph{Loose Policy} provides completely different results. As mentioned previously, the \emph{Loose Policy} uses only the combination of URI-M's Memento-Datetime and the hosting archive as an identifier. The size of the daily observed URI-Ms rises above the total number of observed Memento-Datetimes. This is observed to be an error in the Archeif Web proxy -- the proxy was providing \texttt{00:00:00} as the time portion of the Memento-Datetime for all mementos returned, causing a discrepancy between the \emph{Strict} and \emph{Loose Policies}.

Notice that some Memento-Datetime values are replicated, even though the URI-M is not. This is due to the simultaneous capture of resources by the archives. This produces a discrepancy between the \emph{Strict} and \emph{Loose Policies}.

\section{Evaluation}
\label{caching}
As observed in Section \ref{results}, Memento TimeMaps do not always change for the better. TimeMap cardinality can decrease due to various external influences. The Memento framework proxies currently have a cache with an \emph{infinite} TTL -- TimeMaps are cached on their first occurrence and never automatically replaced. This optimization was critical during the initial deployment of the research prototype Memento Aggregaters. 

We tested a continuum of TTL values to find the best life span of an entity in the cache. A TTL=0 will maximize the freshness of the TimeMaps in the cache, but will also maximize the load on the archives and unnecessarily delay consumer applications. A TTL=$\infty$ will minimize the load on the archives, but also minimize the freshness of the TimeMaps in the cache. Therefore, caching a bad TimeMap is especially bad if we are unfortunate enough to receive a TimeMap with a cardinality of 0, which will remain in the cache until the cache is cleared.

We tested TTL=n, where n=\{0...92\} with 92 effectively equaling $\infty$. We assume the Aggregater  can store an infinite number of TimeMaps and focus only on the possible invalidation of a single TimeMap when its TTL has expired.

\subsection{Caching Policies}
\label{cacheProof}
To determine the most effective cache policy, we tested the \emph{current}, \emph{unconditional}, and \emph{conditional} policies. The \emph{current} caching policy simulates the operation of the Memento Aggregater's current cache with a TTL=$\infty$ to never replace TimeMaps. The \emph{unconditional} policy employs TTL values of 0, $\infty$, and \emph{n}. TTL values set a time limit for cache replacement -- items in a cache are not replaced or removed until a set time limit expires. This policy is used as the baseline measurement since it does not exploit knowledge of the TimeMaps' contents. The \emph{conditional} policy employs TTL values of 0, $\infty$, and \emph{n}, but only replaces TimeMaps in the cache when they have improved according to the \emph{Loose Policy}. The Cases 2-5 are cases that will update an entry in the cache because the TimeMap gains mementos (according to Table \ref{classTable}).

\subsection{Evaluation Measures}
\label{meval}
To measure the impact and quality of the cache replacement policies, the MemDays penalty is defined in Equation \ref{memDaysMeasure}. The MemDays penalty is the sum of the number of mementos that are missed due to a cache hit when there is a better version of the TimeMap available over the time \emph{t} of the experiment.

\begin{equation}
\label{memDaysMeasure}
MemDays = \sum\limits_{1}^t max( \left|TM_{live}\right| - \left|TM_{cache}\right|, 0)
\end{equation}

\begin{figure*}
  \begin{center}
    \includegraphics[width=1\textwidth]{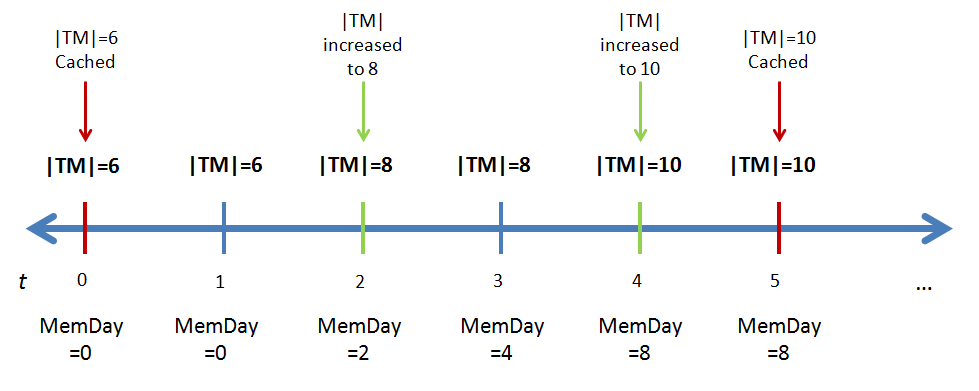}
  \end{center}
  \caption{MemDay calculation example.}
\label{timeline}
\end{figure*}

MemDays  provides a cumulative measure of the detrimental effects of the cache. This metric is the sum of fresh mementos missed for each day those mementos are not available in the cache. An example is provided in Figure \ref{timeline}. A TimeMap with $\left|TM\right|$ = 6 is cached at time \emph{t}=0. At  \emph{t}=1, $\left|TM\right|$=6. Since $\left|TM\right|$ remained consistent, the cumulative MemDay measure is 0. However, $\left|TM\right|$ increases to 8 at time \emph{t}=2. Since the cached $\left|TM\right|$=6, the MemDay measure increases to 2 since 8-6=2. At time \emph{t}=3, the difference between the cached and live $\left|TM\right|$ is still 2, increasing the cumulative MemDay to 4. 

At time \emph{t}=4, the live $\left|TM\right|$=8. The difference between the cached and live $\left|TM\right|$ is now 4, increasing the MemDay by 4 to 8. At \emph{t}=5, the cache is updated with the new TimeMap $\left|TM\right|$=8, and therefore the cached and live $\left|TM\right|$ are equal, adding no additional MemDay penalty.

In this example, Q=2 because the archives contributed to the cached version twice at \emph{t}=0 and \emph{t}=5. The final accumulations of MemDay=8.

The complement to MemDays is the load on the repositories, or amount of queries Q (Equation \ref{q}). This is the sum of the number of times the cache requests a TimeMaps from the archives over the course of \emph{t} days in the experiment, effectively measuring the cache misses due to expired TTL values. Notionally, this is inversely related to the MemDays measure in Equation \ref{memDaysMeasure}. An optimal caching strategy can be determined by minimizing load on the Q and MemDays simultaneously.

\begin{equation}
\label{q}
Q = \sum\limits_{1}^t \text{Cache Misses}
\end{equation}

\subsection{Results}
\label{meval}
At the conclusion of this experiment, we calculated the average number of TimeMap observations in which we observed a monotonically increasing trend. Only 80.2\% of all observations either remained the same or provided an improvement to the existing TimeMap (Figure \ref{optimalttls}). 

To recreate a series of TimeMaps, we used a three-month-long observation period to test the cache replacement policies. We ran a simulation to access the 4,000 TimeMaps with the three caching policies implemented with the \emph{Loose Policy}. The cache size is unlimited for the purposes of this experiment; our goal was only to test the behavior of the replacement policies as they relate to TimeMap change patterns.

\subsubsection{Missed Mementos}
\label{missedmems}
\begin{figure}[h!]
    \includegraphics[width=1\textwidth]{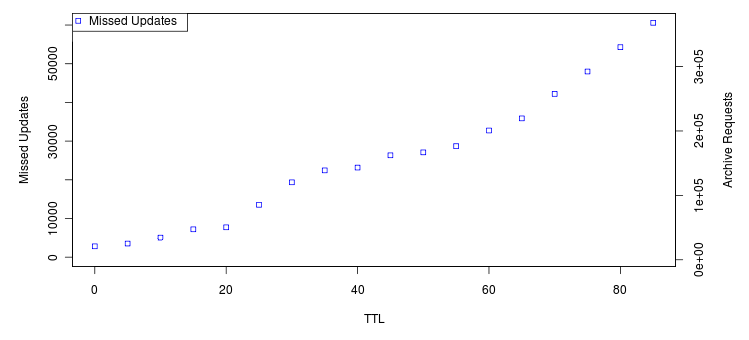}
  \caption{Number of missed cache replacements by TTL value.}
\label{missedupdates}
\end{figure}

As with any cache replacement policy, it is necessary to determine how many updates are missed by utilizing a cache. Since the concept of replacement in this unique cache replacement experiment is limited to those TimeMaps that are \emph{better} (based on the cases in Table \ref{classTable}) than the currently cached version, this experiment counts only those replacements that would result in an improved TimeMap being placed in the cache. The results of this experiment are provided in Figure \ref{missedupdates}. 

\begin{figure}[h!]
    \includegraphics[width=1\textwidth]{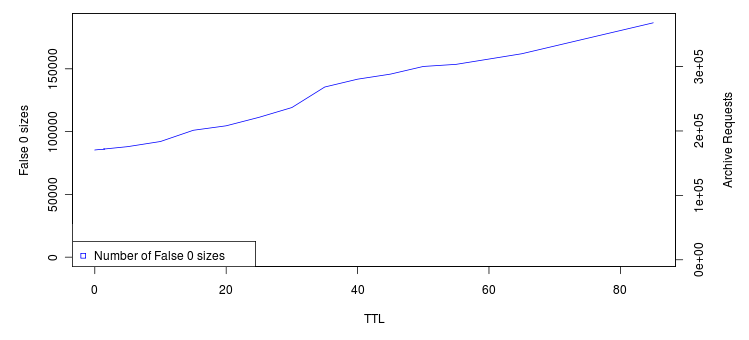}
  \caption{False 0-sized TimeMaps by TTL value for \emph{conditional}.}
\label{false0graphs}
\end{figure}

A primary problem with the current caching algorithm is that TimeMaps with 0 mementos might be cached when TimeMaps with more than 0 mementos exist at another time. False 0-sized TimeMaps (i.e., a 404 response for a TimeMap) are especially detrimental to consumers of TimeMaps since this suggests that a URI-R is not archived when, in fact, it has mementos. This experiment measures the occurrence of these false 0-sized TimeMaps. Additionally, given the rudimentary caching strategies available, a false 0-sized TimeMap should not replace a TimeMap that lists mementos. This is a particularly detrimental occurrence of Case 6 (Table \ref{classTable} -- lost an archive and lost mementos) to consumers of TimeMaps.

As shown in Figure \ref{false0graphs}, the best TTL value to minimize the false 0-sized TimeMaps is 0. However, this induces an unnecessarily large load on the archives. This shows that TimeMaps normally remain constant. However, when they do change, they tend to get worse, as shown in Figure \ref{optimalttls}.
\\
\\
\subsubsection{MemDays versus Q}

\begin{figure*}
  \begin{center}
   \subfigure[MemDay Trade-off by TTL value for \emph{unconditional}.]{\label{memdaypenalties2}\includegraphics[width=1\textwidth]{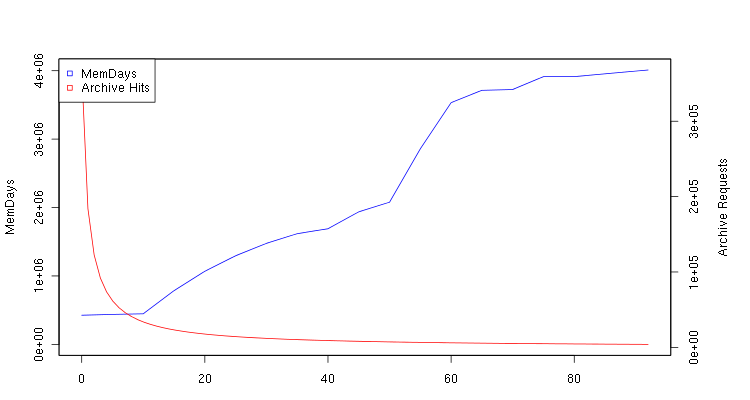}}
   \subfigure[MemDay Trade-off by TTL value for \emph{conditional}.]{\label{memdaypenalties}\includegraphics[width=1\textwidth]{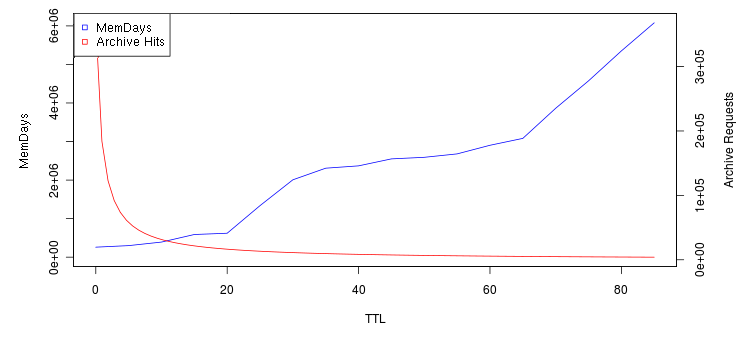}}
  \end{center}
  \caption{Comparison of the \emph{unconditional} and \emph{conditional} policies. MemDays is shown as the blue line, and Q is shown as the red line.}
\label{memdaycomp}
\end{figure*}

The primary measure used in this experiment to determine how well the cache replacement policies are performing is the MemDay penalty (Section \ref{cacheProof}). Using \emph{unconditional}, our experiment shows the MemDay penalty-saved trade-off in Figure \ref{memdaypenalties2}. The MemDays for \emph{conditional} are  shown in Figure \ref{memdaypenalties}. Notice that \emph{conditional} shows improvement over \emph{unconditional} by accumulating fewer MemDays during the simulation. This is because \emph{unconditional} has the potential to cache a TimeMap with $\left|TM\right|$ less than the one in the cache, while \emph{conditional} ensures that the cached TimeMap has the highest $\left|TM\right|$.

As expected, lower TTL values result in fewer missed updates. However, lower TTL values lead to additional queries (Q) to the archives.

In \emph{unconditional}, running a cache that updates TimeMaps at every transaction (TTL=0) results in MemDays=98,312, but Q=368,000 (368,000 requests made to the archives to update the cache). That is, 98,312 mementos are missed because the cache was updated with worse, lower-cardinality TimeMaps. When TTL=$\infty$, MemDays=4,010,406 and \\Q=4,000. TTL=$\infty$ is the \emph{current} policy of the Memento Aggregater.

The optimal TTL value for \emph{unconditional} can be found at the intersection of the Q line (red), and the MemDay line (blue) in Figure \ref{memdaypenalties2}. At this point, MemDays=784,895 due to cache staleness, but Q=92,000. Subsequent data produces a worse trade off between requests to the archives and mementos missed by the users. The data shows that the best TTL value for a TimeMap cache, located at the intersection of the MemDay and Q lines, is 10 days.

In \emph{conditional}, running a cache that updates TimeMaps at every transaction (TTL=0) results in MemDays=0 (or 0 mementos being missed), but Q=368,000 (368,000 requests made to the archives to update the cache). The \emph{conditional} policy improves upon \emph{unconditional} by 98,312 MemDays when TTL=0. Because \emph{conditional} ensures the TimeMap with the highest cardinality remains in the cache, there are no MemDays incurred, as opposed to \emph{unconditional} which blindly assumes improvement to TimeMaps when they change; this is an improper assumption as shown by the change patterns in Table \ref{classTable}. When TTL=$\infty$, MemDays=4,010,406 and Q=4,000, identical to \emph{unconditional}; in both policies, a TTL=$\infty$ caches the first TimeMap available and never replaces it, accumulating MemDays for the duration of the experiment.

The optimal TTL value for \emph{conditional} can be found at the intersection of the Q line (red), and the MemDay line (blue) in Figure \ref{memdaypenalties}. At this point, MemDays=588,368 due to cache staleness (only 7,207 missed TimeMap updates), but Q=23,000. Subsequent data produces a worse trade off between requests to the archives and mementos missed by the users. The data shows that the best TTL value for a TimeMap cache, located at the intersection of the MemDay and Q lines, is 15 days. The optimal TTL for \emph{conditional} improves upon the optimal TTL for \emph{unconditional} by reducing Q by 69,000 and MemDay by 196,527 due to the ability to only update the cache with \emph{better} TimeMaps. This is also intuitively satisfying since TimeMaps change less frequently (every 30 days on average shown in Figure \ref{optimalttls}) than our optimal TTL of 15 days.

\section{Conclusions}
\label{conclusion}
The study outlined in this paper observed 4,000 TimeMaps for URI-Rs over the course of three months and measured their consistency over time. We show that TimeMaps are monotonically increasing only 80.2\% of the time. Due to this evidence, a new \emph{conditional} caching algorithm must be implemented to ensure users have access to the best possible TimeMaps. The current Memento Aggregater operates with a TTL=$\infty$, which produces MemDays=4,010,406 and Q=4,000;  
We have proposed a caching algorithm that implements a TTL value of 15 days and a \emph{conditional} replacement policy in which responses that are smaller than previous responses are not cached. TTL=15 produces MemDays=588,368 and Q=23,000. This shows an improvement on the current Memento Aggregater operation by 3,422,038 MemDays while only making 19,000 more queries to the archives. It improves upon the \emph{unconditional} policy by an observed Q=69,000 and MemDay=196,527.

\section{Acknowledgments}
This work supported in part by the NSF (IIS 1009392) and the Library of 
Congress. We also thank Robert Sanderson (LANL) for the Memento Aggregater research prototype and Ahmed AlSum (ODU) for his assistance with the cache-less Memento Proxies. Memento is a joint project between the Los Alamos National Laboratory Research Library and Old Dominion University

\bibliographystyle{abbrv}

\begin{thebibliography}{}

\end{thebibliography}


\begin{thebibliography}{10}

\bibitem{hmotwia}
S.~Ainsworth, A.~Alsum, H.~SalahEldeen, M.~C. Weigle, and M.~L. Nelson.
\newblock {How much of the Web is archived?}
\newblock In {\em {JCDL '11: Proceedings of the 11th annual international
  ACM/IEEE Joint Conference on Digital Libraries}}, pages 133--136, 2011.

\bibitem{copyright}
{Alyssa N. Knutson}.
\newblock {Proceed With Caution: How Digital Archives Have Been Left in the
  Dark}.
\newblock http://www.btlj.org/data/review/24-437-473.pdf, 2009.

\bibitem{coolUris}
T.~Berners-Lee.
\newblock {Cool URIs don't change}.
\newblock http://www.w3.org/Provider/Style/URI, 1998.

\bibitem{brewington2000keeping}
B.~Brewington, G.~Cybenko, D.~Coll, and N.~Hanover.
\newblock {Keeping up with the changing Web}.
\newblock {\em IEEE Computer}, 33(5):52--58, 2000.

\bibitem{activeCache}
P.~Cao, J.~Zhang, and K.~Beach.
\newblock Active cache: caching dynamic contents on the web.
\newblock In {\em Proceedings of the IFIP International Conference on
  Distributed Systems Platforms and Open Distributed Processing}, pages
  373--388, 1998.

\bibitem{cho2000evolution}
J.~Cho and H.~Garcia-Molina.
\newblock {The evolution of the web and implications for an incremental
  crawler}.
\newblock In {\em Proceedings of the 26th international conference on very
  large data bases}, pages 200--209, 2000.

\bibitem{1267293}
F.~Douglis, A.~Feldmann, B.~Krishnamurthy, and J.~Mogul.
\newblock {Rate of change and other metrics: a live study of the World Wide
  Web}.
\newblock In {\em USITS'97: Proceedings of the USENIX Symposium on Internet
  Technologies and Systems on USENIX Symposium on Internet Technologies and
  Systems}, USITS'97, 1997.

\bibitem{fetterly2004large}
D.~Fetterly, M.~Manasse, M.~Najork, and J.~Wiener.
\newblock {A large-scale study of the evolution of web pages}.
\newblock {\em Software: Practice and Experience}, 34(2):213--237, 2004.

\bibitem{rfc1123}
{Internet Engineering Task Force}.
\newblock {Requirements for Internet Hosts -- Application and Support}, October
  1989.
\newblock http://www.ietf.org/rfc/rfc1123.txt.

\bibitem{cacheDynamic}
A.~Iyengar and J.~Challenger.
\newblock Improving web server performance by caching dynamic data.
\newblock In {\em USITS'97: Proceedings of the USENIX Symposium on Internet
  Technologies and Systems on USENIX Symposium on Internet Technologies and
  Systems}, USITS'97, 1997.

\bibitem{Jaffe:2009:AIA:1534530.1534545}
E.~Jaffe and S.~Kirkpatrick.
\newblock {Architecture of the Internet Archive}.
\newblock In {\em Proceedings of SYSTOR 2009: The Israeli Experimental Systems
  Conference}, SYSTOR '09, 2009.

\bibitem{frank_lazy}
F.~McCown, J.~A. Smith, M.~L. Nelson, and J.~Bollen.
\newblock Lazy preservation: reconstructing websites by crawling the crawlers.
\newblock In {\em WIDM '06: Proceedings of the 8th annual ACM international
  workshop on Web information and data management}, pages 67--74, 2006.


\bibitem{nah}
F.~F. Nah.
\newblock A study on tolerable waiting time: how long are web users willing to
  wait?
\newblock {\em Behaviour \& Information Technology}, (3):153--163.

\bibitem{nelson2011memento}
M.~L. Nelson.
\newblock {Memento-Datetime is not Last-Modified}.
\newblock
  {http://ws-dl.blogspot.com/2010/11/\\2010-11-05-memento-datetime-is-not-last.html},
  2011.

\bibitem{dl_persistence}
M.~L. Nelson and B.~D. Allen.
\newblock Object persistence and availability in digital libraries.
\newblock {\em D-Lib Magazine}, 8(1), January 2002.

\bibitem{ntoulas2004s}
A.~Ntoulas, J.~Cho, and C.~Olston.
\newblock {What's new on the web?: the evolution of the web from a search
  engine perspective}.
\newblock In {\em Proceedings of the 13th international conference on World
  Wide Web}, pages 1--12, 2004.

\bibitem{copyright2}
{Peter B. Hirtle}.
\newblock {Digital Preservation and Copyright}.
\newblock
  http://fairuse.stanford.edu/\\commentary\_and\_analysis/2003\_11\_hirtle.html,
  2012.

\bibitem{sandersonimplementing}
R.~Sanderson, H.~Shankar, S.~Ainsworth, F.~McCown, and S.~Adams.
\newblock {Implementing Time Travel for the Web}.
\newblock {\em Code4Lib Journal}, 13, 2011.

\bibitem{Schwarz2006}
T.~Schwartz, M.~Baker, S.~Bassi, B.~Baumgart, W.~Flagg, C.~van Ingen, K.~Joste,
  M.~Manasse, and M.~Shah.
\newblock {Disk failure investigations at the Internet Archive}.
\newblock {\em 14th NASA Goddard, 23rd IEEE Conference on Mass Storage Systems
  and Technologies}, May 2006.

\bibitem{franktbd}
H.~Tweedy, F.~McCown, and M.~L. Nelson.
\newblock {A Memento Web Browser for iOS}.
\newblock In {\em JCDL '13: Proceedings of the 13th ACM/IEEE-CS Joint
  Conference on Digital Libraries}.

\bibitem{memento:rfc}
H.~{Van de Sompel}, M.~L. Nelson, and R.~Sanderson.
\newblock {HTTP framework for time-based access to resource states -- Memento
  draft-vandesompel-memento-05}.
\newblock https://datatracker.ietf.org/doc/\\draft-vandesompel-memento/, 2012.

\bibitem{nelson:memento:tr}
H.~{Van de Sompel}, M.~L. Nelson, R.~Sanderson, L.~L. Balakireva, S.~Ainsworth,
  and H.~Shankar.
\newblock {Memento: Time Travel for the Web}.
\newblock Technical Report arXiv:0911.1112, 2009.

\bibitem{ldow2010:memento}
H.~{Van de Sompel}, R.~Sanderson, M.~L. Nelson, L.~L. Balakireva, H.~Shankar,
  and S.~Ainsworth.
\newblock {An HTTP-Based Versioning Mechanism for Linked Data}.
\newblock In {\em Proceedings of the Linked Data on the Web Workshop (LDOW
  2010)}, 2010.
\newblock (Also available as arXiv:1003.3661).

\bibitem{webCaching}
J.~Wang.
\newblock A survey of web caching schemes for the {I}nternet.
\newblock {\em SIGCOMM Comput. Commun. Rev.}, 29(5):36--46, Oct. 1999.

\bibitem{mementoviz}
P.~Webster.
\newblock Surfing the web in time: Mementos.
\newblock
  \url{http://britishlibrary.typepad.co.uk/webarchive/2013/01/surfing-the-web-in-time-mementos.html},
  2013.

\bibitem{wessels2004squid}
D.~Wessels.
\newblock {\em Squid: the definitive guide}.
\newblock O'Reilly Media, Incorporated, 2004.

\bibitem{classcache}
H.~Zhu and T.~Yang.
\newblock Class-based cache management for dynamic web content.
\newblock In {\em IEEE INFOCOM}, pages 1215--1224, 2000.

\end{thebibliography}

\end{document}